\documentclass[12pt]{article}

\usepackage{graphicx,amssymb,url,mathrsfs, amsmath}
\usepackage{eucal}
\usepackage{wrapfig}
\usepackage{boxedminipage}
\usepackage{setspace}
\usepackage{subfigure}

\def\a  {\alpha}       \def\b  {\beta}         \def\g  {\gamma}
       \def\d  {\delta}        
        \def\k  {\kappa}
\def\l  {\lambda}             
          \def\s  {\sigma}        
\def\t  {\tau}

\newcommand{\cale}{\mbox{${\cal E}$}}

 \newcommand{\call}{\mbox{${\cal L}$}}
 \newcommand{\caln}{\mbox{${\cal N}$}}

 \newcommand{\scrn}{\mbox{${\mathscr N}$}}

\def\IR{{\hbox{{\rm I}\kern-.2em\hbox{\rm R}}}}
\def\IB{{\hbox{{\rm I}\kern-.2em\hbox{\rm B}}}}
\def\IN{{\hbox{{\rm I}\kern-.2em\hbox{\rm N}}}}
\def\IC{\,\,{\hbox{{\rm I}\kern-.59em\hbox{\bf C}}}}
\def\IZ{{\hbox{{\rm Z}\kern-.4em\hbox{\rm Z}}}}
\def\IP{{\hbox{{\rm I}\kern-.2em\hbox{\rm P}}}}
\def\IH{{\hbox{{\rm I}\kern-.4em\hbox{\rm H}}}}
\def\ID{{\hbox{{\rm I}\kern-.2em\hbox{\rm D}}}}

\def\be{\begin{equation}}
\def\ee{\end{equation}}
\def\ba{\begin{eqnarray}}
\def\ea{\end{eqnarray}}

\def\half{\frac{1}{2}}

\newcommand{\inv}[1]{\frac{1}{#1}}

\def\ra{\rightarrow}

\def\dell{\partial}

\newcommand{\abs}[1]{\left| #1 \right|}

\newcommand{\braket}[2]{\mbox{$\langle #1  | #2 \rangle$}}

\def\Tr{{\rm tr}\,}

\def\det{{\rm det}}

\def\nn{\nonumber}
\def\ea{{\it et al}. }

\newcommand{\gym}{g_{Y\!M}}

\def\De{\textrm{D8}}
\def\DeB{\overline{\textrm{D8}}}

\newcommand{\Ukk}{U_{\rm KK}}
\newcommand{\Mkk}{M_{\rm KK}}

\begin{document}

\begin{titlepage}


\begin{center}
{\large \bf Dense and Hot Holographic QCD:\\
            Finite Baryonic E Field} \\
\vspace{10mm}
  Keun-Young Kim$^a$, Sang-Jin Sin $^{a,b}$  and Ismail Zahed$^a$\\
\vspace{5mm}
$^a$ {\it Department of Physics and Astronomy, SUNY Stony-Brook, NY 11794}\\
$^b$ {\it Department of Physics, BK21 division, Hanyang University, Seoul 133-791, Korea}\\
\vspace{10mm}
\end{center}
\begin{abstract}
We investigate the response of dense and hot holographic QCD
(hQCD) to a static and baryonic electric field $E$ using the
chiral model of Sakai and Sugimoto. Strong fields with
$E>(\sqrt\lambda M_{KK})^2$ free quark pairs, causing the confined
vacuum and matter state to decay. We generalize Schwinger's QED
persistence function to dense hQCD. At high temperature and
density, Ohm's law is derived generalizing a recent result by
Karch and O'Bannon to the chiral case.
\end{abstract}

\end{titlepage}

\renewcommand{\thefootnote}{\arabic{footnote}}
\setcounter{footnote}{0}

\newpage


\section{Introduction}

The AdS/CFT approach~\cite{Maldacena} provides a framework for
discussing large $N_c$ gauge theories at strong coupling
$\lambda=g_{YM}^2N_c$. The model suggested by Sakai and Sugimoto
(SS) \cite{Sakai1} offers a holographic realization of QCD (hQCD)
that has $N_f$ flavors and is chiral. For $N_f\ll N_c$, hQCD is a
gravity dual to $N_f$ $\De$-$\DeB$ branes embedded into a D4
background in 10 dimensions. Supersymmetry is broken by the
Kaluza-Klein (KK) mechanism. The SS model yields a holographic
description of pions, vectors, axials and baryons that is in good
agreement with experiment~\cite{Sakai1, Sakai3, Baryons}. The SS
model at finite temperature has been discussed in the deconfined
phase in~\cite{FiniteT} using a black-hole metric (BH). At finite
baryon density and in the confined phase it has been discussed
in~\cite{Finitemu, KSZ2} using a KK metric. Isospin chemical
potential and glueball have been discussed in~\cite{Others}.

In this paper we would like to continue our investigation of the
model at baryon finite density and temperature but in the
presence of a finite baryonic electric field as recently
discussed by Karch and O'Bannon~\cite{KB07} in a non-chiral model,
as a prelude to understand transport phenomena. In section 2, we
briefly outline the SS model both in the confined KK metric and
deconfined BH metric. In section 3, the DBI action at finite
baryon density is streamlined for both the KK and BH metrics. In
section 4, we discuss Ohm's law in the confined or KK metric.
Above a critical value of the baryon electric field $E>E_c$ the
vacuum and the dense state are unstable against quark pair
creation. In section 5, we show how this pair creation translates
to a vacuum persistence function thereby generalizing Schwinger's
QED result to hQCD both in the vacuum and at finite density. In
section 6, we derive Ohm's law in the BH background, thereby
extending a recent result by Karch and Bannon~\cite{KB07} to the
chiral case. The vacuum instability is dwarfed by thermal pair
creation in the incoherent statistical averaging with a treshold
value for the baryonic electric field starting at zero. Our
conclusions are in section 7.

\section{Sakai-Sugimoto model}

In this section we summarize the Sakai-Sugimoto model
(D4/$\De$-$\DeB$ set up) for notation and completeness. For a
thorough presentation we refer \cite{Sakai1} for zero temperature
and \cite{FiniteT} for finite temperature.

At zero temperature, the confined KK  metric,  dilaton $\phi$,
and the 3-form RR field $C_3$ in $N_c$ D4-branes background are
given by
\begin{eqnarray}
&&ds^2=\left(\frac{U}{R}\right)^{3/2} \left(-dt^2 + \d_{ij}dx^i
dx^j+f(U)(dx^4) ^2\right) +\left(\frac{R}{U}\right)^{3/2}
\left(\frac{dU^2}{f(U)}+U^2 d\Omega_4^2\right)\  ,
\nn\\
&&~~~~e^\phi= g_s \left(\frac{U}{R}\right)^{3/4}, ~~F_4\equiv
dC_3=\frac{2\pi N_c}{V_4}\epsilon_4 \ , ~~~f(U)\equiv
1-\frac{\Ukk^3}{U^3} \ , \label{LowT}
\end{eqnarray}
where  $x^i= x^{1,2,3}$ and $U (\geq U_{KK})$ and $\Omega_4$ are
the radial coordinate and four angle variables in the
$x^{5,6,7,8,9}$ direction. $R^3 \equiv \pi g_s N_c l_s^3$, with
$g_s$ and $l_s$ the string coupling and length respectively.
$V_4=8\pi^2/3$ is the volume of the unit $S^4$ and $\epsilon_4$ is
the corresponding volume form. To avoid a conical singularity at
$U=\Ukk$ the period of $\d\t$ of the compactified $\t$ direction
is set to $\d\t = \frac{4\pi}{3}\frac{R^{3/2}}{\Ukk^{1/2}}$.
The Kaluza-Klein mass is
\begin{eqnarray}
\Mkk \equiv
 \frac{2\pi}{\d\t} = \frac{3}{2}
\frac{\Ukk^{1/2}}{R^{3/2}} \ .\nn
\end{eqnarray}
The parameters $R$, $\Ukk$, and $g_s$ may be
expressed in terms of $\Mkk$, $\l( = \gym N_c)$, and $l_s$ as
\begin{eqnarray}
R^3 = \half \frac{\l  l_s^2}{\Mkk}\ , \quad \Ukk = \frac{2}{9} \l
\Mkk l_s^2 \ , \quad g_s = \frac{1}{2\pi}\frac{\l}{\Mkk N_c l_s}\,\,.
\ \nn
\end{eqnarray}

At finite temperature, there are two possibilities. One is the
same as the zero temperature apart from the fact the time
direction is Euclidean and compactified with a circumference $\b
= 1/T$. It corresponds to the confined phase which is the low
temperature regime(KK). The other corresponds to deconfined phase
of the high temperature. Its geometry contains the black hole and
the pertinent BH background is
\begin{eqnarray}
&&ds^2=\left(\frac{U}{R}\right)^{3/2} \left(f(U)dt_E^2  +
\d_{ij}dx^{i}dx^{j}+ (dx^4)^2\right)
+\left(\frac{R}{U}\right)^{3/2}
\left(\frac{dU^2}{f(U)}+U^2 d\Omega_4^2\right)\ \nn , \label{HighT} \\
&&e^\phi= g_s \left(\frac{U}{R}\right)^{3/4}, ~~F_4\equiv
dC_3=\frac{2\pi N_c}{V_4}\epsilon_4  , \quad f(U)\equiv
1-\frac{U_T^3}{U^3} \ ,
\end{eqnarray}
The $t_E$-direction must be periodic
\begin{eqnarray}
\d t_E = \frac{4\pi}{3}\left( \frac{R^3}{U_T}\right)^{1/2} \equiv
\b = \frac{1}{T}
\end{eqnarray}
to avoid a conical singularity. The $x^4$-direction is also
periodic but it has an arbitrary periodicity.

Now, consider $N_f$ probe D8-branes in the $N_c$ D4-branes
background. With $U(N_f)$ gauge field $A_M$ on the D8-branes, the
effective action consists of the DBI action and the Chern-Simons
action:
\begin{eqnarray}
S_{\De}&=& S^{DBI}_{\De} + S^{CS}_{\De}\ , \label{Action.0} \nn \\
S^{DBI}_{\De}&=& -T_8 \int d^9 x \ e^{-\phi}\ \Tr
\sqrt{-\det(g_{MN}+2\pi\alpha' F_{MN})} \ , \label{DBI0}\\
S^{CS}_{\De}&=&\frac{1}{48\pi^3} \int_{D8} C_3 \Tr F^3 \label{CS0}
\ .
\end{eqnarray}
where $T_8 = 1/ ((2\pi)^8 l_s^9)$, the tension of the D8-brane,
$F_{MN}=\partial_M A_N -\partial_N A_M -i \left[ A_M , A_N
\right]$ ($M,N = 0,1,\cdots,8$), and $g_{MN}$ is the induced
metric on  D8-branes. The expressions are written for the
Minkowskian metric.

\section{DBI action}

The induced metric on the D8 branes from the gravity background
(\ref{LowT}) and (\ref{HighT}) may be written as
\begin{eqnarray}
ds_{\mathrm{D8}}^2 &\equiv& g_{tt} dt^2 +g_{xx}
\d_{ij}dx^{i}dx^{j} + {g_{UU}}\, dU^2 + g_{SS}
d\Omega_4^2 \  \label{D8metricabs} \\
&\equiv&  \a\left(\frac{U}{R}\right)^{3/2} dt^2 +
\left(\frac{U}{R}\right)^{3/2}\d_{ij}dx^{i}dx^{j} +
\left(\frac{R}{U}\right)^{3/2} \g\, dU^2 +
\left(\frac{R}{U}\right)^{3/2}U^2d\Omega_4^2 \ , \label{D8metric}
\end{eqnarray}
where for the KK background
\begin{eqnarray}
\a \ \ra\ -1 \  , \quad \g \ \ra\ \inv{f(U)} + \left(\frac{\dell
x^4}{\dell U}\right)^2 \left(\frac{U}{R} \right)^3 f(U) \ , \quad
f(U)\ \ra\ 1-\left(\frac{\Ukk}{U}\right)^3 \ ,  \label{KK}
\end{eqnarray}
and for the BH background
\begin{eqnarray}
\a \ra f(U) \  , \quad \g \ra \inv{f(U)} + \left(\frac{\dell
x^4}{\dell U}\right)^2 \left(\frac{U}{R} \right)^3 \ , \quad
f(U)\ \ra\ 1-\left(\frac{U_T}{U}\right)^3 \ . \label{BH}
\end{eqnarray}
The embedding information is  encoded only in $\gamma$ and thereby
$g_{UU}$. We will use the abstract metric notations
(\ref{D8metricabs})  to treat the confined  (\ref{KK}) and deconfined
(\ref{BH}) coherently in formal  evaluation here. In
the next section we will plug in the specific embedding and metric
form.

To accommodate a static baryonic electric field on $\De$ branes
both in vacuum and matter, we follow~\cite{KB07}  to define
\begin{eqnarray}
  A_t = A_t(U) \ , \quad A_x = -Et + h_x(U) \ . \label{KarchGauge}
\end{eqnarray}
With the induced metric (\ref{D8metricabs}) and the guage fields
(\ref{KarchGauge}) the DBI action (\ref{DBI0}) is written as
\begin{eqnarray}
&&S_{\mathrm{DBI}} \equiv \int d^4x  dU \call_{\mathrm{DBI}} \nn \\
&&= - \caln \int dU \ e^{-\phi} g_{SS}^2 g_{xx}
\sqrt{|g_{tt}|g_{xx}g_{UU} - (2\pi\a')^2\left(g_{xx}(A_t')^2 +
g_{UU}(\dot{A}_{x})^2 - |g_{tt}|(A_x')^2 \right) }
\label{DBI}
\end{eqnarray}
where $\caln \equiv (2 N_f) T_8 V_4$. $2N_f$ comes from the fact
that we consider $N_f$ branes and anti-branes and $V_4(=8/3\pi^2)$
is the volume of the unit $S^4$ which is due to the trivial
integral over $S^4$. $'$ is the derivative with respect to $U$ and
$\ \dot{}\ $ is the derivative with respect to $t$. Since
(\ref{DBI}) is purely kinetic, the conjugate momenta $D$ and $B$
are conserved. Specifically,
\begin{eqnarray}
&&D \equiv \frac{\dell\call_{\mathrm{DBI}}}{\dell A_t'} =
e^{-\phi} g_{SS}^2 g_{xx} \frac{- \caln (2 \pi \a')^2 g_{xx}
A_t'}{\sqrt{|g_{tt}| g_{xx} g_{UU} - (2 \pi \a')^2 \left( g_{xx}
A_t'^2 + g_{UU} E^2 - |g_{tt}| h_x'^2 \right )}} \label{D} \\
&&B \equiv \frac{\dell\call_{\mathrm{DBI}}}{\dell A_x'} =
e^{-\phi} g_{SS}^2g_{xx} \frac{\caln (2 \pi \a')^2 |g_{tt}|
h_x'}{\sqrt{|g_{tt}| g_{xx} g_{UU} - (2 \pi \a')^2 \left( g_{xx}
A_t'^2 + g_{UU} E^2 - |g_{tt}| h_x'^2 \right )}} \label{B}
\end{eqnarray}
By rewriting $A_t'$ and $h_x'$  in terms of $B$, $D$ and $E$, we have
\begin{eqnarray}
&& g_{xx} A_t'(U)^2 = \frac{1}{(2 \pi \a')^2} |g_{tt}| D^2
\frac{g_{UU} (|g_{tt}| g_{xx} - (2 \pi \a')^2 E^2)}{\caln^2 (2 \pi
\a')^2 |g_{tt}| g_{xx}^3 e^{-2\phi} g_{SS}^4   + |g_{tt}| D^2 -
g_{xx} B^2} \\
&& |g_{tt}| h_x'(U)^2 = \frac{1}{(2 \pi \a')^2} g_{xx} B^2
\frac{g_{UU} (|g_{tt}| g_{xx} - (2 \pi \a')^2 E^2)}{\caln^2 (2
\pi \a')^2 |g_{tt}| g_{xx}^3  e^{-2\phi} g_{SS}^4 + |g_{tt}| D^2 -
g_{xx} B^2}
\end{eqnarray}
The DBI action reduces to
\begin{eqnarray}
S_{\mathrm{DBI}} = -\caln \int d^4x dU \left[ e^{-2\phi}g_{SS}^4
g_{xx}^{5/2} |g_{tt}|^{1/2} g_{UU}^{1/2} \right]
\sqrt{\frac{(|g_{tt}| g_{xx} - (2 \pi \a')^2 E^2)}{|g_{tt}|
g_{xx}^3 e^{-2\phi}g_{SS}^4 + \frac{|g_{tt}| D^2 - g_{xx} B^2 }{
\caln^2 (2 \pi \a')^2 }}} \label{dbi}
\end{eqnarray}
Notice that $g_{tt}, g_{xx}, g_{SS}$ have nothing to do with the
D8 branes embedding. They carry information of D4 branes. Only
$g_{UU}$ carries information of the $x^4(U)$. It is positive for
all $U$. Thus the factors outside the square root are real
for all $U$. In contrast, the argument of square root may
change the sign for varying $U$. As we will discuss below, this
change in sign is the signal of a ground state instability  or
decay  for large $E$ fields.

%

\section{Ohm's law: KK}

This decay is captured by a non-linear form of
Ohm's law. For that, it is useful to change variable
\begin{eqnarray}
U = U_0 (1 + Z^2)^{1/3} \ , \label{UtoZ}
\end{eqnarray}
where $U_0$ is the coordinate of the tip of $\De$-$\DeB$ branes'
cigar-shaped configuration, which is different from $\Ukk$ in
general.  The range of $Z$ is $(0,\infty)$ contrary to $U$ whose
range is $(U_0,\infty)$. Also this range can be extended to
$(-\infty,\infty)$ if we consider $\DeB$ branes $(-\infty,0)$
together with $\De$ branes $(0, \infty)$ in a natural way. It
enables us to deal with the ADHM instanton solution in
$\mathbb{R}^4$~\cite{Sakai3}. It also makes the parity property
of the meson fields explicit \cite{Sakai1}. For completeness, we
note the following useful relations
\begin{eqnarray}
K \equiv 1+Z^2 \ , \quad U = U_0K^{1/3} \ , \quad dU = \frac{2U_0
}{3}\frac{Z}{K^{2/3}} dZ\ , \quad f =
1-\left(\frac{\Ukk}{U_0}\right)^3\inv{K} \ .
\end{eqnarray}
From here on and for simplicity, we follow Sakai and
Sugimoto~\cite{Sakai1} and choose $U_0 = \Ukk$. The DBI action
then simplifies to
\begin{eqnarray}
  S_{\mathrm{DBI}} = -a\int d^4x dZ K^{1/6} \sqrt{\frac{K- \frac{b}{\Mkk^2} E^2}{1+\frac{D^2-B^2}{a^2
  b} K^{-5/3} }} \ ,
\end{eqnarray}
where
\begin{eqnarray}
  a \equiv \frac{N_c N_f \l^3 \Mkk^4}{3^9 \pi^5}\ , \qquad b
  \equiv \frac{3^6 \pi^2}{4 \l^2 \Mkk^2} \ .
\end{eqnarray}
In dense hQCD baryons are sourced by BPST instantons in bulk with
a size of order $1/\sqrt{\lambda}$. They are point-like at
$\lambda\rightarrow\infty$. Thus the DBI action and the matter
sources read
\begin{eqnarray}
  \call_\mathrm{tot} = \call_{\mathrm{DBI}}+{n}_B \d(Z) A_t(Z) + \widetilde{n}_B v_x \d (Z)
  A_x(t,Z)\ ,
\label{SOURCE}
\end{eqnarray}
where $n_B$ is the baryon - anti baryon density and
$\widetilde{n}_B$ is baryon + anti baryon density. The first
source contribution is that of static BPST instantons at $Z=0$ as
initially discussed in~\cite{KSZ2}. The second term is their
corresponding current with a velocity $v_x\sim 1/{\lambda N_c}$
with a baryon mass $M_B\sim N_c\lambda M_{KK}$. Note that we have
renormalized the $A_\mu$ field here by $1/N_c$ and identified the
baryon chemical potential as $A_\mu(\infty)=\mu_B -
m_B$~\cite{KSZ2}.

The equations of motion are
\begin{eqnarray}
   D' =  n_B \  \d(Z) \ , \qquad B' = \widetilde{n}_B v_x \d(Z) \ .
\end{eqnarray}
Thus
\begin{eqnarray}
  D = \half n_B \mathrm{sgn}(Z)  , \qquad B = \half \widetilde{n}_B v_x \mathrm{sgn}(Z) \
  ,
\end{eqnarray}
where $\mathrm{sgn}(Z)$ reflects the symmetry of $\De$ and $\DeB$
branes (chirality). We note that the conserved momenta $D,B$ are
odd functions of $Z$ since the baryonic field $A_\mu$ is an even
function of $Z$.

For a finite baryonic electric field $E$, the current contribution
in (\ref{SOURCE}) is seen to increase linearly with time in the action.
This is expected since the static electric field pumps energy in
the system. For times $t\sim M_B\sim
N_c\lambda$ the present stationary (time-independent) surface analysis is
flawed. This notwithstanding, the action variation with respect
to $A_t$ yields
\begin{eqnarray}
  \d_{A_t} S_\mathrm{tot} &=& \int dZ \left[ \frac{\d \call}{\d (\dell_Z A_t)} \dell_Z (\d A_t) + n_B \d(Z) \d A_t(Z)
  \right]\nn \\
  &=& \int dZ  \left(\half n_B \mathrm{sgn}(Z)  \dell_Z (\d A_t) \right) + n_B  \d
  A_t(0) \nn \\
  &=& n_B \d A_t(\infty) \ .
\end{eqnarray}
where we used the on-shell condition and  $A_t(\infty) =
A_t(-\infty)=\mu_B - m_B$. Note that the contribution from the
source term is cancelled by the boundary contribution of the DBI
action at $Z=0$. As a result the on-shell action may be
considered as a functional of $A_t(\infty)$ only and we may set
$A_t(0)=0$. Similarly for $A_x(t,0) = 0$,
\begin{eqnarray}
\d_{A_t} S_\mathrm{tot}= \widetilde{n}_B \,v_x\d A_x(t,\infty) \ .
\end{eqnarray}
The former is the charge, while the latter is the current. At
finite density $S_\mathrm{tot}$ plays the role of the grand
potential. Thus
\begin{eqnarray}
   \bar{S} = -a\int d^4x dZ K^{1/6} \sqrt{\frac{K- \frac{b}{\Mkk^2} E^2}{1+\frac{n_B^2- \widetilde{n}_B^2 v_x^2 } {4 a^2
  b}K^{-5/3} }} ,
\label{ACTIONX1}
\end{eqnarray}
on shell. For $v_x = E = 0$ this result is consistent with our
previous result i.e. Eq.(30) in~\cite{KSZ2} which is indeed the
grand potential.

For $0 < E\leq E_c \equiv \frac{\Mkk}{\sqrt{b}}$, $J_x
 (=\widetilde{n}_Bv_x)$ is bounded,
\begin{eqnarray}
J_x < \sqrt{4a^2 b + n_B^2} \ ,
\end{eqnarray}
for $\bar{S}$ to be real. For $E>E_c$, the numerator of
(\ref{ACTIONX1}) flips sign at
\begin{eqnarray}
K_* = \frac{b}{\Mkk^2}E^2 \ , \qquad Z_* = \pm \sqrt{  b E^2 -
1}\,\ .
\end{eqnarray}
We demand that this flip is compensated by the denominator for
arbitrary $v_x$. Using $Z_*$ in the denominator we get
\begin{eqnarray}
J_x^2 &=& 4 a^2 b K_*^{5/3} + n_B^2 \nn \\  &=&
\inv{2^{10/3}3^2\pi^{14/3}}N_f^2N_c^2\left(\frac{\l}{\Mkk}\right)^{2/3}
E^{10/3} + n_B^2 \ \theta(E)\ .  \label{Jx2}
\end{eqnarray}
In the unstable vacuum, the ensuing Ohmic's conductivity is
\begin{eqnarray}
  \s \equiv \frac{J_x }{E} = \frac{1}{2^{5/3}3\pi^{7/3}}
  N_cN_f\left(\frac{\l}{\Mkk}\right)^{1/3} E^{2/3} \ .
\end{eqnarray}
This pair conductivity follows from quark pairs and not from
baryon pairs as it scales with $N_cN_f$. $E_c$ is strong enough
to cause deconfinement of quark pairs. For $n_B\neq 0$ the second
contribution in (\ref{Jx2}) is that of the baryons and anti
baryons moving  under the action of the {\it strong} electric
field, with $\Delta v\sim Et/M_B\sim t/N_c$. Note that for $E=0$,
the minimum of (\ref{ACTIONX1}) is for $v_x = 0$.

For $E>E_c$ both the vacuum with $n_B=0$ and the dense baryonic
state with $n_B\neq 0$ are unstable against pair creation of
quark-antiquark states as opposed to baryon-antibaryon states.
This is clearly seen from the threshold value $E_c$
\begin{equation}
E_c = \frac{\Mkk}{\sqrt{b}}= \frac{2}{27\pi}\Mkk^2
 \lambda  = \frac{54 \pi M_B^2}{\l N_c^2} \,\,,
\end{equation}
with $M_B = 8\pi^2\k \Mkk$ and $\k = \frac{\l
N_c}{216\pi^3}$~\cite{Sakai3} which is much smaller than $M_B^2$.
The baryonic electric field is strong enough to pair create
quarks with {\it constituent masses} of order
$\sqrt{\lambda}M_{KK}$ \footnote{It is interesting to note that
in the BH background the thermal shifts of heavy quarks is
$\pi\sqrt{\lambda} T/2$ with $T$ in the unconfined phase being
the analogue of $M_{KK}$ in the confined phase.}.

\section{Persistence Probability}

The cold and dense states described by hQCD above are
unstable for $E>E_c$, meaning that they decay
to multiparticle states that are likely time-dependent.
Following Schwinger, we will characterize this decay
through its persistence probability
\begin{eqnarray}
  \abs{\braket{0_+}{0_-}}^2 = e^{-2\mathrm{Im}\bar{S} },
\end{eqnarray}
where $\mathrm{Im}\bar{S}$ is the imaginary part of the action
$\bar{S}$ (\ref{ACTIONX1}). For finite $n_B$ and $ v_x  = 0$,
the action $\bar{S}$ reads
\begin{eqnarray}
  \bar{S} = - a \int d^4x dZ (1+Z^2)^{1/6}\left[\sqrt{\frac{Z^2 + 1 - \cale^2}{1 + \scrn^2 (1+Z^2)^{-5/3} }} -
  \sqrt{\frac{Z^2+1}{1 + \scrn^2 (1+Z^2)^{-5/3} }}\right],
\end{eqnarray}
with $\cale^2 \equiv \frac{b}{\Mkk^2}E^2, \ \scrn^2 \equiv
\frac{n_B^2}{4a^2b}\, $ and after regularizing the action by
subtracting the $\cale = 0$ contribution. $E_c$ corresponds to
$\cale_c = 1$. For $\cale \le 1$ the action $\bar{S}$ is always
real, but for $\cale
> 1$ the action develops an imaginary part from the integration
interval $(-Z_c,Z_c)$, where $Z_c \equiv \sqrt{\cale^2-1}$. Thus
\begin{eqnarray}
  \mathrm{Im}\bar{S} &=& \pm a \int d^4x  \int_{-Z_c}^{Z_c} dZ (1+Z^2)^{1/6}
  \sqrt{\frac{Z^2 + 1 - \cale^2}{1 + \scrn^2 (1+Z^2)^{-5/3} }} \theta(\cale-1)\,\,.
\end{eqnarray}

For $\scrn = 0$ the integrals unwind analytically
\begin{eqnarray}
  \mathrm{Im}\bar{S}
  = \pm a \pi \int d^4x \left[(\cale^2-1)\
  _2F_1\left(-\inv{6},\half,2,1-\cale^2\right)\theta(\cale-1)\right].
\end{eqnarray}
where $_2F_1$ is the hypergeometric function and has the
asymptotic behaviour as follows.
\begin{eqnarray}
&&_2F_1\left(-\inv{6},\half,2,1-\cale^2\right) \nn \\
&& \qquad \sim 1+\inv{12}(\cale-1) + \inv{144}(\cale-1)^2 + \cdots    \qquad \qquad \qquad (\cale \sim 1)\nn \\
&& \qquad  \sim
\frac{\Gamma(2/3)}{\sqrt{\pi}\Gamma(13/6)}\cale^{1/3} +
\frac{2\Gamma(-2/3)}{\sqrt{\pi}\Gamma(-1/6)}\inv{\cale} + \cdots
\qquad \qquad (\cale \gg 1)
\end{eqnarray}
The persistence function is then
\begin{eqnarray}
  \abs{\braket{0_+}{0_-}}^2 &=& e^{- a' (\varepsilon^2-1)\
  _2F_1\left(-\inv{6},\half,2,1-\varepsilon^2\right)\theta(\varepsilon-1)
  },  \nn \\
  &=&  1 \qquad \qquad   \qquad \qquad \qquad \qquad  (\cale \le 1) \nn \\
    &~&  e^{- a' [2 (\varepsilon -1) + 1.17(\varepsilon-1)^2 +
    \cdots]}
    \qquad    \quad  (\cale \sim 1)
  \nn \\
    & ~&  e^{- a' [0.71\ \varepsilon^{7/3} + 0.70 \ \varepsilon +
    \cdots]}
    \qquad \qquad    (\cale \gg 1)
\end{eqnarray}
with $a' \equiv a \pi \int d^4x = \frac{N_c N_f \l^3 \Mkk^4}{3^9
\pi^4} \int d^4x $, after chosing the negative sign for decay.

\section{Ohm's law: BH}

Since the vacuum decay under large $E$'s so does the coherent
finite baryonic state. But what about the finite temperature problem?
As finite temperature involves a statistical ensemble averaging,
we may suggest that the unstable ground state is statistically irrelevant
and proceed to analyse the effects of a baryonic field on the excited states
(unstable by fiat) in the ensemble average. This will be checked a posteriori
below.

In the BH background there are two possible gravitational
configurations: 1/ a U-shaped (chirally broken phase) and 2/ a
parallell-shape (chirally symmetric phase). The former yields $U$
bounded from below by $U_0$. The combination $g_{tt}g_{xx}$ has a
positive minimum so the numerator is always positive for
sufficiently small E. The nature of the transition which is
suggestive of a metal-insulator transition~\cite{Meyer} will be
discussed elsewhere.

For high enough temperature the stable configuration is not the
U-shaped configuration but the parallel configuration which is
connected to the black hole. i.e. $\frac{dx^4}{dU}= 0$. Our
intial instanton sources have now drowned into the BH horizon.
So the ensuing analysis is the same as in the D3/D7 model~\cite{KB07},
with the general formula of the conductivity for  Dq/Dp given ((5.7) in
\cite{KB07}). Here and for completeness, we compute the
conductivity for the parallel $\De$-$\DeB$ branes set up in the BH
background.

We only need to consider the positivity  condition for the argument of
square root as before. As $U
\ra U_T$ \textit{both} the numerator and denominator are
\textit{negative} since $g_{tt} \ra 0$. As $U \ra \infty$
\textit{both} the numerator and denominator are
\textit{positive}. So by choosing B,D,E we can choose the
numerator and denominator in (\ref{dbi}) to flip sign for the
same value $U=U_*$~\cite{KB07}. For the numerator
\begin{eqnarray}
|g_{tt}|g_{xx} \Big|_{U=U_*} = (2\pi\a')^2 E^2 \nn \\
\Rightarrow U_* = (U_T^3 + R^3(2\pi\a'E)^2)^{1/3}\ .
\end{eqnarray}
Inserting this value of $U_*$ in the denominator yields the
induced current
\begin{eqnarray}
J_x^2 &=&
\left(\caln^2(2\pi\a')^2|g_{tt}|g_{xx}^2e^{-2\phi}g_{SS}^4
+ \frac{|g_{tt}|}{g_{xx}}\, J_t^2 \right) \Big|_{U=U_*} \nn \\
&=& \left(\frac{\caln^2(2\pi\a')^4 R^6}{g_s^2}\left(U_T^3 +
R^3(2\pi\a'E)^2\right)^{2/3} +
\frac{(2\pi\a')^2}{\frac{U_T^3}{R^3} + (2\pi\a'E)^2}\, J_t^2
\right) E^2\,\,  \ ,
\end{eqnarray}
where $J_x = B$ and $  J_t = D $ ($=n_B$) are now defined as
in~\cite{KB07}. Setting $U_T = \frac{16\pi^2}{9}T^2R^3$, $\l=g_s
N_c$ yield the Ohmic conductivity for the chiral SS model
\begin{eqnarray}
  \s = \frac{ J_x }{E} = \sqrt{ \left(\frac{4 l_s N_f N_c\l T^2 }{27}\right)^2  (1 + e^2)^{2/3} +
  \frac{d^2}{1+e^2}}\  ,  \label{sigmahigh}
\end{eqnarray}
where
\begin{eqnarray}
e \equiv \frac{3^3 E}{2^5 \pi^3 T^3 \l l_s} \ ,  \qquad d \equiv
\frac{3^3  J_t }{2^5 \pi^3 T^3 \l l_s} \ ,
\end{eqnarray}
which is consistent with the result in~\cite{KB07} for massless
but non-chiral quarks. The induced thermal current sets in for any
$E\geq 0$ (large or small) with a conductivity $\sigma$ of order
${N_cN_f\lambda T^2l_s}$ at high temperature, which involves only
thermal pairs with zero treshold for $E$. It dwarfs the induced
vacuum pairs by a factor of ${\lambda}^{2/3}$. The unstable vacuum
state is statistically irrelevant. This is not the case at $T=0$
and/or very large baryonic densities.

\section{Conclusions}
We have extended our recent holographic analysis of the SS model
at finite density, to the case of finite temperature and finite
baryonic electric field. For $E>E_c$ the stationary SS ground
state breaks down by quark pair creation. This phenomenon
permeates both the cold and hot states of hQCD. The vacuum
persistence probability is derived, generalizing Schwinger's QED
result to hQCD. At finite temperature, the baryonic electric
field yields a thermal conductivity at finite temperature and
density that is a direct generalization of Karch and Bannon's
Ohm's law in the chiral model. We have argued that the vacuum
instability is statistically irrelevant in hot hQCD.

\section{Acknowledgments}
The work of KYK and IZ was supported in part by US-DOE grants
DE-FG02-88ER40388 and DE-FG03-97ER4014. The work of SJS was
supported by KOSEF Grant R01-2007-000-10214-0 and the SRC Program
of the KOSEF through the CQUEST with grant number R11-2005-021.

\section{Note added}

While typing our results, a recent analysis appeared
in~\cite{Bergman} that addresses similar issues in the model with
cusp surfaces at $L\neq \pi$. Our results are all for cuspless
surfaces with $L=\pi$ for the KK background, and the parallel
or deconfined configuration for the BH background.


\begin{thebibliography}{99}

\bibitem{Maldacena}
  J.~M.~Maldacena,
  {\sl The large N limit of superconformal field theories and supergravity},
  {\sl Adv.\ Theor.\ Math.\ Phys.\  {\bf 2}, 231 (1998)},
  {\sl Int.\ J.\ Theor.\ Phys.\  {\bf 38}, 1113 (1999)}
  [arXiv:hep-th/9711200]; For review, see: \\
  O.~Aharony, S.~S.~Gubser, J.~M.~Maldacena, H.~Ooguri and Y.~Oz,
  {\sl Large N field theories, string theory and gravity},
  Phys.\ Rept.\  {\bf 323}, 183 (2000)
  [arXiv:hep-th/9905111].

\bibitem{Sakai1}
  T.~Sakai and S.~Sugimoto,
  ``Low energy hadron physics in holographic QCD,''
  Prog.\ Theor.\ Phys.\  {\bf 113}, 843 (2005)
  [arXiv:hep-th/0412141]; \\
  T.~Sakai and S.~Sugimoto,
  ``More on a holographic dual of QCD,''
  Prog.\ Theor.\ Phys.\  {\bf 114}, 1083 (2006)
  [arXiv:hep-th/0507073].

\bibitem{Sakai3}
  H.~Hata, T.~Sakai, S.~Sugimoto and S.~Yamato,
  ``Baryons from instantons in holographic QCD,''
  arXiv:hep-th/0701280.

\bibitem{Baryons}
  K.~Nawa, H.~Suganuma and T.~Kojo,
  ``Baryons in Holographic QCD,''
  Phys.\ Rev.\  D {\bf 75}, 086003 (2007)
  [arXiv:hep-th/0612187]; \\
  D.~K.~Hong, M.~Rho, H.~U.~Yee and P.~Yi,
  ``Chiral dynamics of baryons from string theory,''
  arXiv:hep-th/0701276;\\
    D.~K.~Hong, M.~Rho, H.~U.~Yee and P.~Yi,
  ``Dynamics of Baryons from String Theory and Vector Dominance,''
  arXiv:0705.2632 [hep-th]; \\
  D.~K.~Hong, M.~Rho, H.~U.~Yee and P.~Yi,
  ``Nucleon Form Factors and Hidden Symmetry in Holographic QCD,''
  arXiv:0710.4615 [hep-ph].



\bibitem{FiniteT}
  O.~Aharony, J.~Sonnenschein and S.~Yankielowicz,
  ``A holographic model of deconfinement and chiral symmetry restoration,''
  Annals Phys.\  {\bf 322}, 1420 (2007)
  [arXiv:hep-th/0604161]; \\
  A.~Parnachev and D.~A.~Sahakyan,
  ``Chiral phase transition from string theory,''
  Phys.\ Rev.\ Lett.\  {\bf 97}, 111601 (2006)
  [arXiv:hep-th/0604173]; \\
  K.~Peeters, J.~Sonnenschein and M.~Zamaklar,
  ``Holographic melting and related properties of mesons in a quark gluon
  plasma,''
  Phys.\ Rev.\  D {\bf 74}, 106008 (2006)
  [arXiv:hep-th/0606195].

\bibitem{Finitemu}
  K.~Y.~Kim, S.~J.~Sin and I.~Zahed,
  ``Dense hadronic matter in holographic QCD,''
  arXiv:hep-th/0608046; \\
  N.~Horigome and Y.~Tanii,
  ``Holographic chiral phase transition with chemical potential,''
  JHEP {\bf 0701}, 072 (2007)
  [arXiv:hep-th/0608198]; \\
  O.~Bergman, G.~Lifschytz and M.~Lippert,
  ``Holographic Nuclear Physics,''
  arXiv:0708.0326 [hep-th]; \\
  M.~Rozali, H.~H.~Shieh, M.~Van Raamsdonk and J.~Wu,
  ``Cold Nuclear Matter In Holographic QCD,''
  arXiv:0708.1322 [hep-th];

\bibitem{KSZ2}
  K.~Y.~Kim, S.~J.~Sin and I.~Zahed,
  ``The Chiral Model of Sakai-Sugimoto at Finite Baryon Density,''
  JHEP {\bf 0801}, 002 (2008)
  [arXiv:0708.1469 [hep-th]].
 \bibitem{KSZ3}
  K.~Y.~Kim, S.~J.~Sin and I.~Zahed,
  ``Dense Holographic QCD in the Wigner-Seitz Approximation,''
  arXiv:0712.1582 [hep-th].


\bibitem{Others}
  A.~Parnachev,
  ``Holographic QCD with Isospin Chemical Potential,''
  arXiv:0708.3170 [hep-th]; \\
  O.~Aharony, K.~Peeters, J.~Sonnenschein and M.~Zamaklar,
  ``Rho meson condensation at finite isospin chemical potential in a
  holographic model for QCD,''
  arXiv:0709.3948 [hep-th]; \\
  K.~Hashimoto, C.~I.~Tan and S.~Terashima,
  ``Glueball Decay in Holographic QCD,''
  arXiv:0709.2208 [hep-th].


\bibitem{KB07}
  A.~Karch and A.~O'Bannon,
  ``Metallic AdS/CFT,''
  JHEP {\bf 0709}, 024 (2007)
  [arXiv:0705.3870 [hep-th]].




\bibitem{Meyer}
  J.~Erdmenger, R.~Meyer and J.~P.~Shock,
  ``AdS/CFT with Flavour in Electric and Magnetic Kalb-Ramond Fields,''
  JHEP {\bf 0712}, 091 (2007)
  [arXiv:0709.1551 [hep-th]].


\bibitem{Bergman}
  O.~Bergman, G.~Lifschytz and M.~Lippert,
  ``Response of Holographic QCD to Electric and Magnetic Fields,''
  arXiv:0802.3720 [hep-th].





\end{thebibliography}
\end{document}